\documentclass[twocolumn]{aa}

\usepackage{graphicx,txfonts,natbib,verbatim}
\bibpunct{(}{)}{;}{a}{}{,}

\begin{document}

\title{Positions and sizes of X-ray solar flare sources}

\author{E. P. Kontar \and N. L. S. Jeffrey}

\offprints{E.P. Kontar \email{eduard@astro.gla.ac.uk}}

\institute{Department of Physics \& Astronomy, University of Glasgow, G12 8QQ, United Kingdom}

\date{Received ; Accepted }

\abstract{}{To investigate the positions and source sizes of X-ray sources
taking into account Compton backscattering (albedo).}
{Using a Monte Carlo simulation of X-ray photon transport including photo-electric absorption
and Compton scattering, we calculate the apparent source sizes
and positions of X-ray sources at the solar disk for various source sizes, spectral
indices and directivities of the primary source.}
{We show that the albedo effect will alter the true source positions and substantially increase
the measured source sizes. The source positions are shifted up to $\sim 0.5''$ radially towards
the disk centre and 5 arcsecond source sizes can be two times larger even for an isotropic
source (minimum albedo effect) at 1~Mm above the photosphere. X-ray sources therefore
should have minimum observed sizes, thus FWHM source size (2.35 times second-moment)
will be as large as $\sim 7''$ in the 20-50 keV range for a disk-centered
point source at a height of 1~Mm ($\sim 1.4''$) above the photosphere. The source size
and position change is the largest for flatter primary
X-ray spectra, stronger downward anisotropy, for sources closer to the solar
disk centre, and between the energies of 30 and 50 keV.}
{Albedo should be taken into account when X-ray footpoint positions, footpoint motions
or source sizes from e.g. RHESSI or Yohkoh data are interpreted, and suggest that
footpoint sources should be larger in X-rays than in optical or EUV ranges.}


\titlerunning{Positions and sizes of X-ray solar flare sources}

\authorrunning{Kontar and Jeffrey}

\maketitle

\section*{Introduction}
Hard X-ray (HXR) emission produced via collisional bremsstrahlung
from solar flares by non-thermal electrons is the primary
diagnostic tool of electron acceleration and transport.
The spectral and spatial distributions of HXR
sources provide us with vital clues to improve current understanding
of the underlying physics involved in
energetic electron acceleration and transport.
While recent \citep[Hard X-ray Telescope (HXT) on Yohkoh,][]{Kosugi_etal1991},
and modern \citep[Reuven Ramaty High Energy Solar Spectroscopic Imager,][]{Lin_etal2002},
solar HXR telescopes have provided
superb X-ray image resolution,  indirect imaging
using either pairs of occultation
grids on Yohkoh or rotating modulating collimators
on RHESSI have (i) limited dynamic range and (ii) often provided
inadequate spatial resolution exceeding the size
of e.g. EUV footpoints or vertical extend of the chromosphere.
Thus RHESSI image resolution for the majority of solar
flares is limited to about $7''$, while the
solar chromosphere is only about $3''$ thick.
Nevertheless, unprecedented spatial measurements
can and have been achieved using the {\it moments}
of X-ray distributions.
The total flux ({\it zeroth moment}) from individual sources
in various energy ranges allows imaging
spectroscopy \citep[e.g.][]{kruckerlin2002,emslie2003,BattagliaBenz2007,Saint-Hilaire_etal2008}.
The measurements of the X-ray source positions ({\it first moments})
pin-point source locations with $1''$ or better accuracy
and allow us to infer the chromospheric density structure
\citep{aschwanden_etal2002,Liu_etal2006,Kontar_etal08}.
The motions of HXR footpoint locations have been used to infer
the reconnection rate in solar flares \citep{FletcherHudson2002,Krucker_etal2003,fivian2009}.
Using X-ray visibilities \citep{Hurford_etal2002,Schmahl_etal2007}
\citet{Kontar_etal08} have measured not only the positions
but the HXR footpoint sizes ({\it second moment}) at various energies
and heights and found that HXR sources decrease with energy
and consequently with height above the photosphere.
\citet{Xu_etal2008} have measured coronal
sources to infer acceleration region sizes. HXR images can also be
inverted \citep[e.g.][]{Brown_etal2006} to find the spatial electron
distributions and hence the locations of electron centroids \citep[e.g.][]{Prato_etal2009}.

Since the solar atmosphere above HXR sources is optically thin,
X-rays are often directly related to the emitting electrons.
However, the photons emitted downwards, toward the denser layers
of the atmosphere interact with free or bound electrons and
can also be scattered toward the observer \citep{Tomblin72,Santangelo_etal73}.
Photons back-scattered and emerging back from the dense solar atmosphere
to the observer create the albedo X-ray photons. Even for an isotropic X-ray source
(the minimum albedo), the albedo flux can account
for up to 40\% of the detected flux in the range
between 30 and 50 keV \citep{BaiRamaty1978,ZhangHuang2004,Kontar_etal2006,Kasparova_etal2007}. Therefore,
all X-ray sources at the solar disk are viewed as a combination
of both the primary and backscattered fluxes. Accounting for the albedo effect
is important for all X-ray solar observations, which can only
view disk sources as a combination of the primary photon flux
and the backscattered photon flux. The backscattered component taints the
primary source properties such as electron angular, energy, and spatial distributions.
Albedo changes the shape of the spatially integrated X-ray
spectrum, which is flattened at lower energies up to around 20-30 keV and can even
produce artificial spectral features in observed spectra \citep{KontarDickson2008},
while at higher energies above around 70 keV, the the spectrum is steeper
than the emitted (primary) spectrum.
\citet{Kontar_etal2006} have developed and implemented albedo correction
for spectral X-ray RHESSI analysis using Green's
functions approximations by \citet{MagdziarzZdziarski95}.
Since the reflected X-rays come from a rather large area (albedo patch),
the surface brightness of the albedo patch at the solar surface is rather
low \citep{BaiRamaty1978}. This fact explains the difficulty in directly
imaging the albedo patch \citep{SchmahlHurford2002},
but highlights the importance of the inclusion of albedo for understanding
the measurements  of the source positions and sizes (first and second moments),
the quantities which are integrated over the full area of the source.

In this Letter, using Monte Carlo simulations of X-ray photon transport
we demonstrate how the observed positions and source sizes are
affected by the albedo effect for various anisotropies,
primary source sizes and primary source spectra and show that
on-disk HXR sources should have
energy-dependent minimum observed sizes.


\section*{Spatial characteristics of the primary, backscattered and observed
X-ray distributions}
\begin{figure*}
\centering
\includegraphics[width=180mm]{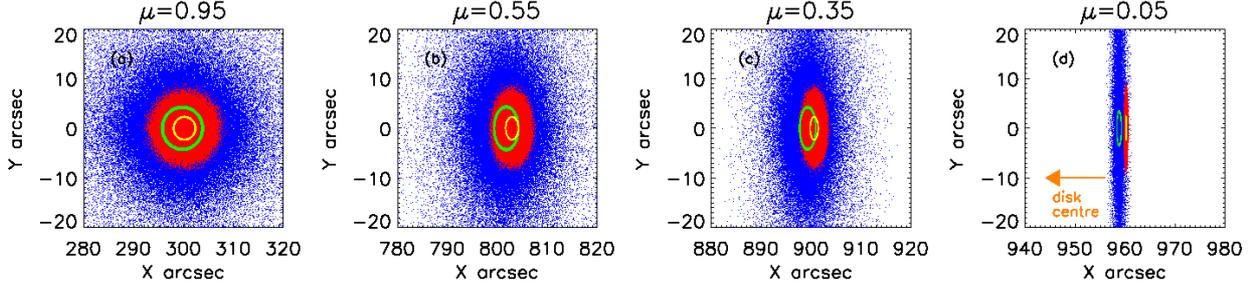}
\caption{The X-ray scatter distributions of the primary photons (red dots)
and the Compton back-scattered photons (blue dots) for a primary source
at $h=1.0$~Mm with $d=1.5$~Mm (FWHM$\sim 4.9''$)
between 20 and 50 keV for four viewing angles given by $\mu$.
The yellow and green ellipses show the FWHM sizes for the primary
and combined sources respectively.}
\label{fig1}
\end{figure*}

\begin{figure*}
\centering
\includegraphics[width=180mm]{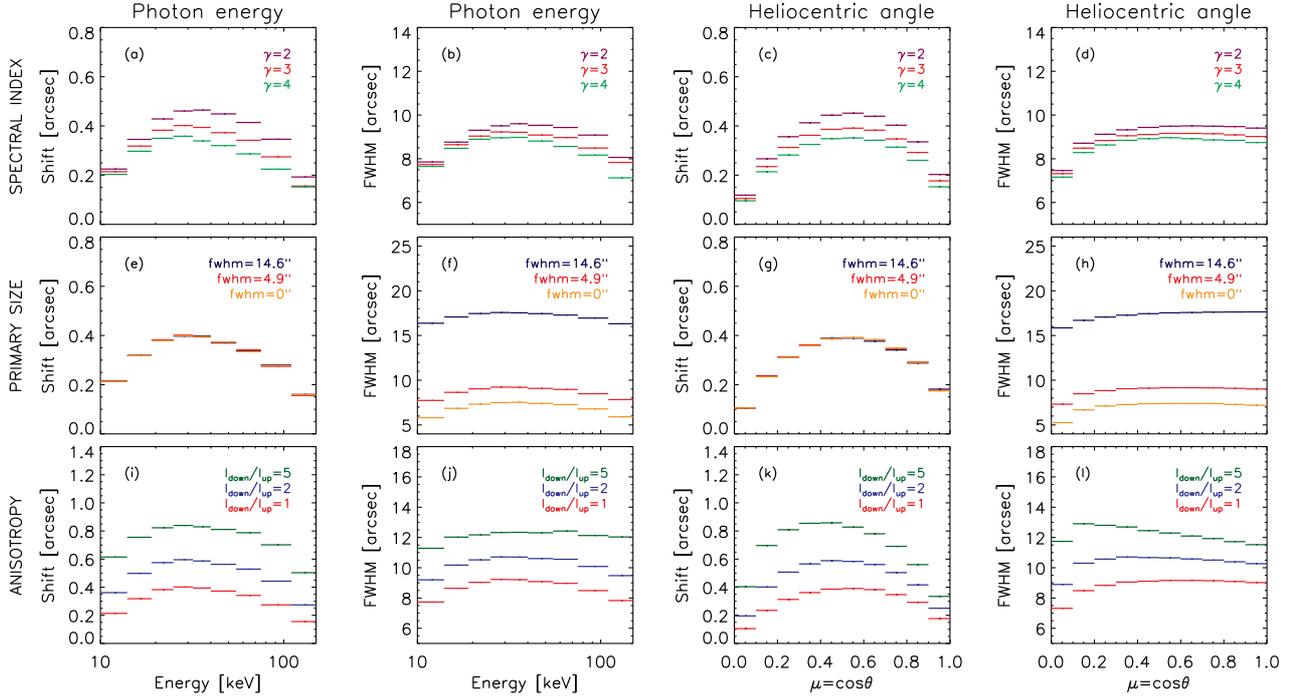}
\caption{{\it Spectral index dependency} (panels a-d):
The source position shift is in the radial direction
due to albedo (source $\mu=0.55$) and source size FWHM is in the perpendicular to radial
direction for various spectral indices $\gamma$ for an isotropic source
with FWHM$\sim 4.9''$: green- $\gamma =4$; red - $\gamma =3$; purple -$\gamma =2$.
{\it Primary source size dependency} (panels e-h): Isotropic
primary source with $\gamma=3$: orange - point source, red - FWHM$\sim4.9''$,
blue - FWHM$\sim14.6''$. {\it Anisotropy dependency} (ratio of downward
to upward directed fluxes) (panels i-l): Simulations are for the primary
source with FWHM$\sim4.9''$ and spectral index $\gamma=3$:
red - anisotropy=1 (isotropic), blue - anisotropy=2, green - anisotropy=5.
Graphs as a function of energy are for $\mu=0.55$ and the graphs as a function of $\mu$
are for energies between 20 and 50 keV.}
\label{fig2}
\end{figure*}

The backscattered flux and albedo effect are studied using a Monte Carlo simulation
starting with a hundred million photons per run.  An unpolarized X-ray source was modeled
in space with a 2-dimensional circular Gaussian $\sim \exp [-x^2/(2d^2)-y^2/(2d^2)]$
with width $d$, placed at the height $h=1$~Mm above the photosphere [The photosphere is defined
here as a layer with Hydrogen number density $1.16\times10^{17}$cm$^{-3}$ \citep[][]{vernazza1981}].
This is the typical hard X-ray source height found
in footpoints \citep{aschwanden_etal2002,Kontar_etal08}.
The energy spectrum for photons has a power law $I(\epsilon)\sim \epsilon^{-\gamma}$
with a spectral index of $\gamma$, for energies between $3$ keV and $300$ keV,
typical for RHESSI. The code accounts
for the curvature of the Sun and the photons are assumed to move freely
until they reach the photospheric density
at a height $z_{\odot}=\sqrt{R_{\odot}^{2}-x^2-y^2}-R_{\odot}$,
where $R_{\odot}=6.96\times 10^{10}$~cm is the solar radius.
Below this level photons can be either scattered or photo-electrically
absorbed. Similar to previous MC simulations \citep{BaiRamaty1978,MagdziarzZdziarski95},
the Klein-Nishina cross-section for unpolarized X-ray radiation was used
\begin{equation}
\frac{d\sigma_c}{d\Omega}(\epsilon_0,\theta_s)=\frac{1}{2}r_0{^2}\left(\left(\frac{\epsilon}{\epsilon_0}\right)^3
+\frac{\epsilon}{\epsilon_0}-\left(\frac{\epsilon}{\epsilon_0}\right)^2\sin{^2}\theta_s\right),
\label{eq:Klein-Nishina}
\end{equation}
where $\epsilon_0$ is the initial photon energy,
$\epsilon$ is the new photon energy, $\theta_s$ is the angle between the initial
and new photon direction and $r_0=2.82\times10^{-13}$cm is the classical electron
radius. After a scattering, the new photon energy is just given by $\epsilon=\epsilon_0/(1+\frac{\epsilon_0}{mc^{2}}(1-\cos\theta_s))$.
The absorption of X-ray photons, which is the dominant process
below $\sim 10 $~keV was modeled using modern solar
photospheric abundances \citep{Asplund_etal2009}
and cross-sections \citep{Henke_etal1982,Balucinska-ChurchMcCammon1992}
for the most important elements H, He, C, N, O, Ne, Na, Mg, Al, Si, S, Cl, Ar, Ca, Cr, Fe and Ni.
For X-ray energies $> 10 $~keV, photoelectric absorption was approximated
as $\sigma_a(\epsilon_0)\sim \epsilon_0^{-3}$ \citep{MagdziarzZdziarski95}.
To account for elements with more electrons than Hydrogen, e.g. Helium,
Carbon etc, Equation (\ref{eq:Klein-Nishina}) was multiplied by 1.18.
Our simulations differ from previous simulations
\cite[e.g.][]{BaiRamaty1978,MagdziarzZdziarski95} because of newer
abundances and the inclusion of the curvature of the Sun.
The escaping photons are accumulated to create the brightness
distribution $I(x,y)$ over a given energy and solid angle.
The total primary or reflected flux is then
just an integral over the corresponding area $\int I(x,y)dxdy$.
Fig.~\ref{fig1}a shows the primary and escaping photon brightness
distributions for a source
located at the disk centre. Similar to the previous results \citep{BaiRamaty1978} we see that
for a compact primary source of size $d=1.5h$, the back-scattered (albedo) photons
are reflected from an area much larger than the primary source.
The reflected photons change the spatial distribution of the observed photons
and produce a halo around the primary source. Importantly, even a primary point source
will be seen as a source of finite size (Fig. \ref{fig2}).
The brightness distribution of a large primary source of $d=4.5h$ is less
influenced by the reflected photons but nevertheless the source will
look larger than it actually is.

Using solar disk centered coordinates, the centroid position
of the source ($\bar{x}$, $\bar{y}$) can be found by calculating
the first normalized moment of the distribution (mean)
\begin{equation}
\bar{x}=\frac{\int_{-\infty}^{\infty}xI(x,y)dxdy}{\int_{-\infty}^{\infty}I(x,y)dxdy}\;\;\; \mbox{,} \;\;\;
\bar{y}=\frac{\int_{-\infty}^{\infty}yI(x,y)dxdy}{\int_{-\infty}^{\infty}I(x,y)dxdy}
\label{eq:xy_c}
\end{equation}
and the normalized variance of the distribution (second moment),
\begin{equation}
\sigma^2_x=\frac{\int_{-\infty}^{\infty}(x-\bar{x})^2I(x,y)dxdy}{\int_{-\infty}^{\infty}I(x,y)dxdy}
\; \mbox{,} \;
\sigma^2_y=\frac{\int_{-\infty}^{\infty}(y-\bar{y})^2I(x,y)dxdy}{\int_{-\infty}^{\infty}I(x,y)dxdy}.
\end{equation}
Hereafter, following RHESSI measurements \citep[][]{Kontar_etal08,dennis2009,Prato_etal2009}
we will use source sizes in terms of FWHM (Full Width Half Maximum),
$FWHM_{x,y}=2\sqrt{2\ln{2}}\sigma_{x,y}$.

The scattered X-ray flux depends on the cosine of heliocentric angle of the
source ($\mu \equiv\cos (\theta)$) or equivalently on the position of the source
at the solar disk, $\mu =\sqrt{1-(x^2+y^2)/R_{\odot}^2}$.
A circular X-ray source located above the centre of the disk will
produce a circular albedo patch  (Fig. \ref{fig1}a).
 Naturally,
the location of the HXR source and albedo patch will coincide
at the disk centre, so albedo will not change
the source position. However, the albedo will make the source larger than it is
actually is (Fig. \ref{fig1}a). The albedo contribution becomes asymmetric
if the source is located away from the disk centre at a given heliocentric
angle $\theta$ (Fig. \ref{fig1}b-d).
Due to the spherical symmetry of the Sun,
there are two distinct directions: radial - along the line connecting the centre of the Sun
and the X-ray source $r$, and perpendicular to the radial $r_\perp$.
There is no change in centroid position in the $r_\perp$-direction for a spherically
symmetric primary source. In the $r$-direction, the albedo causes a centroid
shift towards the disk centre that rises from $\sim 0$
at $\mu=1.0$ and peaks shortly before falling to $\sim 0$
again at $\mu=0.0$. Fig. \ref{fig1} also shows how the source size varies
in $r_{\perp}$ direction, with the FWHM of the source generally
decreasing at lower $\mu$. In the radial direction, the FWHM of the total
and primary sources decreases close to linear due to a simple projection effect.
The detailed 3D structure of the source is required before any physically meaningful predictions
can be made concerning the change in source size in the radial direction, and this is beyond the scope of the paper.
Therefore, we consider the source sizes in the $r_{\perp}$ direction and the source
position in the radial direction rather than along the East-West and
South-North directions. Similar to the spatially integrated albedo \citep{Kontar_etal2006},
the shift in centroid position and the growth of the source are also energy
and $\mu$ dependent. In the following, we consider the position
and source size changes for various a) spectra of the primary source, b) primary source size,
and c) X-ray  directivity (the ratio of downward to upward emitted photons) separately.
The results are summarized in Fig. \ref{fig2}.

{\bf Spectral index} (Fig. \ref{fig2}{a-d}) -
Similar to the spectral results,
the albedo contribution from a smaller spectral index
produces the largest shift in position and the larger
total source size (Fig. \ref{fig2}a-d). An isotropic source
of FWHM$\sim 4.9''$ with the smallest modeled spectral
index of $\gamma =2$ produces the greatest
shift of $\sim 0.5''$ at $\mu=0.5-0.6$
and $\sim 30$ keV. This spectral index also produces
the largest source size and has a FWHM$\sim 9.5''$ at $\mu=1.0$,
compared with the other spectral indices of $\gamma=3,4$ modeled.

{\bf Primary source size} (Fig. \ref{fig2}{e-h}) -
For a fixed spectral index of $\gamma=3$, all primary
source sizes produce the same shift in centroid position.
The maximum shift in position occurs at $\mu=0.5-0.6$ and $\sim30$ keV
for all sources (Fig. \ref{fig2}e,g).
Although the FWHM of the total source grows with increasing
primary size, it is observed that the relative size
of the total to the primary source is smaller for a larger
primary source. This indicates that a larger primary source should
have a smaller relative size increase due to albedo.
Even an initial point source produces a total source
with a FWHM peaking around $7''$ (Fig. \ref{fig2}f,h).

{\bf Anisotropy} (Fig. \ref{fig2}{i-l}) - The shift in centroid position
is larger for a higher initial downward anisotropy
(the ratio of downward flux to upward flux)
for all $\mu$ and energies (Fig. \ref{fig2}i,k).
All shifts follow the general trend and tend towards
zero at the centre ($\mu=1.0$) and the limb ($\mu=0.0$).
Using $\gamma=3$ and a primary source of FWHM$\sim 4.9''$,
a directivity of 5 produces a peak difference of $\sim 0.9''$ and even
an isotropic source produces a peak difference of $\sim 0.4''$.
The shift in source position peaks near $\mu=0.4-0.6$
and $\sim30$ keV for a downward
anisotropy of 2 and an isotropic source, but the shift peaks
at a lower $\mu=0.4-0.5$ for a downward directivity of 5.
The stronger downward beaming of the primary
source also leads to larger apparent source sizes for all $\mu$
and energies (Fig. \ref{fig2}j,l).
It should be observed that the total FWHM produced
for a directivity of 5 peaks at $\mu\sim0.15$ (Fig. \ref{fig2}p) giving an apparent
FWHM$\sim 13''$. Since the fraction of reflected photons
reduces with $\mu$ the FWHM in perpendicular direction
can be expected to slowly decrease from disk centre to limb,
but the FWHM actually increases, peaks at $\mu\sim 0.15$ and only
then starts to decrease. This effect is due to the angular dependence
of the Compton cross-section. The cross-section is anisotropic
and peaks at $90^o$, which allows larger number of photons
to scatter into an observer direction for near the limb flare.
It is this anisotropy in the scattering of the photons
that causes the FWHM to peak at an angle smaller than $\mu$=1.0.
The high photon flux from a downward directivity of 5 allows the observation
of this effect most clearly (Fig. \ref{fig2}l).

\section*{Discussion and conclusions}

The results of the simulations show that albedo can substantially
affect the precise position and source size measurements of X-ray sources.
Therefore, the effect of albedo should always be (probably with the exception
of limb/occulted flares) considered  when the sizes or positions of X-ray
sources are analyzed. The albedo displacement of the source position is
radially directed towards the disk centre and depends on anisotropy of X-ray radiation,
the X-ray source size and the spectral index of the primary source. Similar to total
reflected flux, the displacement of HXR source position is energy dependent.
The largest displacement can be observed in the range between
30-50 keV at $\mu \sim 0.5$ (heliocentric angle $\sim 60^o$).
The shift in centroid position in this energy range is $0.1-0.5''$
for an isotropic (minimum albedo) source $1.4''$ above the photosphere
and this can be up $\sim 0.9''$ for downward beaming with factor of $5$. Because
of the albedo, X-ray source sizes will be
energy dependent, larger in the perpendicular to radial
direction, and elliptical even for a spherically symmetric primary source.
In the perpendicular to radial direction, the largest growth in source size appears
for sources close to the solar disk centre, in the energy range between 30-50 keV, where
albedo is the strongest. Thus, an isotropic
primary source with FWHM $\sim 4.9''$ at $1.4''$ above the photosphere will
have an apparent FWHM size of $\sim 9''$ in the energy range
20-50 keV for sources in the wide range of heliocentric angles from $0^o$ to
$\sim 80^o$.

The simulations demonstrate that X-ray sources will have a minimum size.
An isotropic point source at $1.0$~Mm above the photosphere will be measured
by RHESSI as a source with a FWHM size of $\sim 7''$ across.
This result can explain larger X-ray footpoint
sizes than EUV or optical ones \citep[e.g.][]{Kasparova_etal2005}.
\citet{dennis2009} reported that the average semi-minor axis
of 18 double source flares is about $4''$, while a few of the
X-ray source sizes were found to be consistent with line sources along
the flare ribbons. While the quantitative
comparison with the RHESSI observations requires additional work,
we note that zero sizes are either the artifacts of the algorithms
used or are due to very low source heights.

The energy dependent character of albedo predicts that the source size
as measured by RHESSI should grow with energy from $10$~keV up to $\sim 30$~keV.
Considering a large primary source of $14.6''$ across, e.g. a flaring loop,
we find that the source will grow up to $\sim 18''$ at $\sim 30$ keV.
Noteworthy, \citet{Xu_etal2008} have found that coronal source sizes
are growing with energy along both the field lines and across.
While the field line increase along the lines could be an indicator
of electron transport or of the acceleration region size, the cross-field increase
remains unexplained, but is consistent with the growth of the source size
due to the albedo. We note that the spatial changes of X-ray sources due to
albedo have a great diagnostic potential for purely known anisotropy
of energetic electrons.

\begin{acknowledgements}

The authors are indebt to G. Hurford for insightful comments.
EPK work is supported by a STFC rolling grant, STFC Advanced Fellowship
and the Leverhulme Trust grant. NLSJ work was supported by The Nuffield Foundation
and Cormack Bequest, Royal Society Edinburgh. The work has benefited
from the international team grant from ISSI, Bern, Switzerland.

\end{acknowledgements}

\bibliographystyle{aa}
\bibliography{references1}

\end{document}